\documentclass[submitting]{nst}

\usepackage{subfigure,dcolumn}
\usepackage{epstopdf}
\usepackage{mhchem}
\usepackage{upgreek}

\begin{document}

\title{Transition edge sensor based detector: from X-ray to $\gamma$-ray\mbox{}}
\thanks{Project supported by the National major scientific research instrument development project (Grant No.11927805), NSFC Young Scientists Fund (Grant No. 12005134), Shanghai-XFEL Beamline Project (SBP) (Grant No. 31011505505885920161A2101001), Shanghai Municipal Science and Technology Major Project(Grant No.2017SHZDZX02),Shanghai Pujiang Program(Grant No.20PJ1410900)}

\author{Shuo Zhang}
\email[Corresponding author, ]{shuozhang@shanghaitech.edu.cn}
\affiliation{Center for Transformative Science, ShanghaiTech University, ShangHai, 201210, China}

\author{Jing-Kai Xia}
\affiliation{Center for Transformative Science, ShanghaiTech University, ShangHai, 201210, China}

\author{Tao Sun}
\affiliation{Shanghai Institute of Microsystem and Information Technology, Chinese Academy of Sciences, ShangHai, 200050, China}

\author{Wen-Tao Wu}
\affiliation{Shanghai Institute of Microsystem and Information Technology, Chinese Academy of Sciences, ShangHai, 200050, China}

\author{Bing-Jun Wu}
\affiliation{Shanghai Institute of Microsystem and Information Technology, Chinese Academy of Sciences, ShangHai, 200050, China}

\author{Yong-Liang Wang}
\affiliation{Shanghai Institute of Microsystem and Information Technology, Chinese Academy of Sciences, ShangHai, 200050, China}

\author{Robin Cantor}
\affiliation{STAR Cryoelectronics, 25-A Bisbee Court, Santa Fe, NM, 87508-1338}

\author{Ke Han}
\affiliation{INPAC and Department of Physics and Astronomy, Shanghai Jiao Tong University, Shanghai Laboratory for Particle Physics and Cosmology, Shanghai, 200240, China}

\author{Xiao-Peng Zhou}
\affiliation{School of Physics, Beihang University, Beijing 100191, China}

\author{Hao-Ran Liu}
\affiliation{Division of Ionizing Radiation, National Institute of Metrology, NIM, No.18 Bei San Huan Dong Lu, Chaoyang Dist, Beijing, 100029, China}

\author{Fu-You Fan}
\affiliation{Division of Ionizing Radiation, National Institute of Metrology, NIM, No.18 Bei San Huan Dong Lu, Chaoyang Dist, Beijing, 100029, China}

\author{Si-Ming Guo}
\affiliation{Division of Ionizing Radiation, National Institute of Metrology, NIM, No.18 Bei San Huan Dong Lu, Chaoyang Dist, Beijing, 100029, China}

\author{Jun-Cheng Liang}
\affiliation{Division of Ionizing Radiation, National Institute of Metrology, NIM, No.18 Bei San Huan Dong Lu, Chaoyang Dist, Beijing, 100029, China}

\author{De-Hong Li}
\affiliation{Division of Ionizing Radiation, National Institute of Metrology, NIM, No.18 Bei San Huan Dong Lu, Chaoyang Dist, Beijing, 100029, China}

\author{Yan-Ru Song}
\affiliation{Center for Transformative Science, ShanghaiTech University, ShangHai, 201210, China}

\author{Xu-Dong Ju}
\affiliation{Center for Transformative Science, ShanghaiTech University, ShangHai, 201210, China}

\author{Qiang Fu}
\affiliation{Center for Transformative Science, ShanghaiTech University, ShangHai, 201210, China}

\author{Zhi Liu}
\affiliation{Center for Transformative Science, ShanghaiTech University, ShangHai, 201210, China}

\begin{abstract}
The Transition Edge Sensor(TES) is extremely sensitive to the change of temperature, combined with the high-Z metal of a certain thickness, it can realize high energy resolution measurement of particles such as X-ray. X-ray with energies below 10 keV have very weak penetrating ability, so only a few microns thick of gold or bismuth can guarantee a quantum efficiency higher than 70\%. Therefore, the entire structure of the TES X-ray detector for this energy range can be realized with the microfabrication process. However, for X-ray or $\gamma$-ray from 10 keV to 200 keV, sub-millimeter absorber layers are required, which cannot be realized by microfabrication process. This paper first briefly introduces a set of TES X-ray detectors and their auxiliary systems, then focus on the introduction of the TES $\gamma$-ray detector, with absorber based on an sub-millimeter lead-tin alloy sphere. The detector has a quantum efficiency above 70\% near 100 keV, and an energy resolution of about 161.5eV@59.5keV.
\end{abstract}


\maketitle

\begin{figure*}[!htb]
\includegraphics[width=.7\hsize]{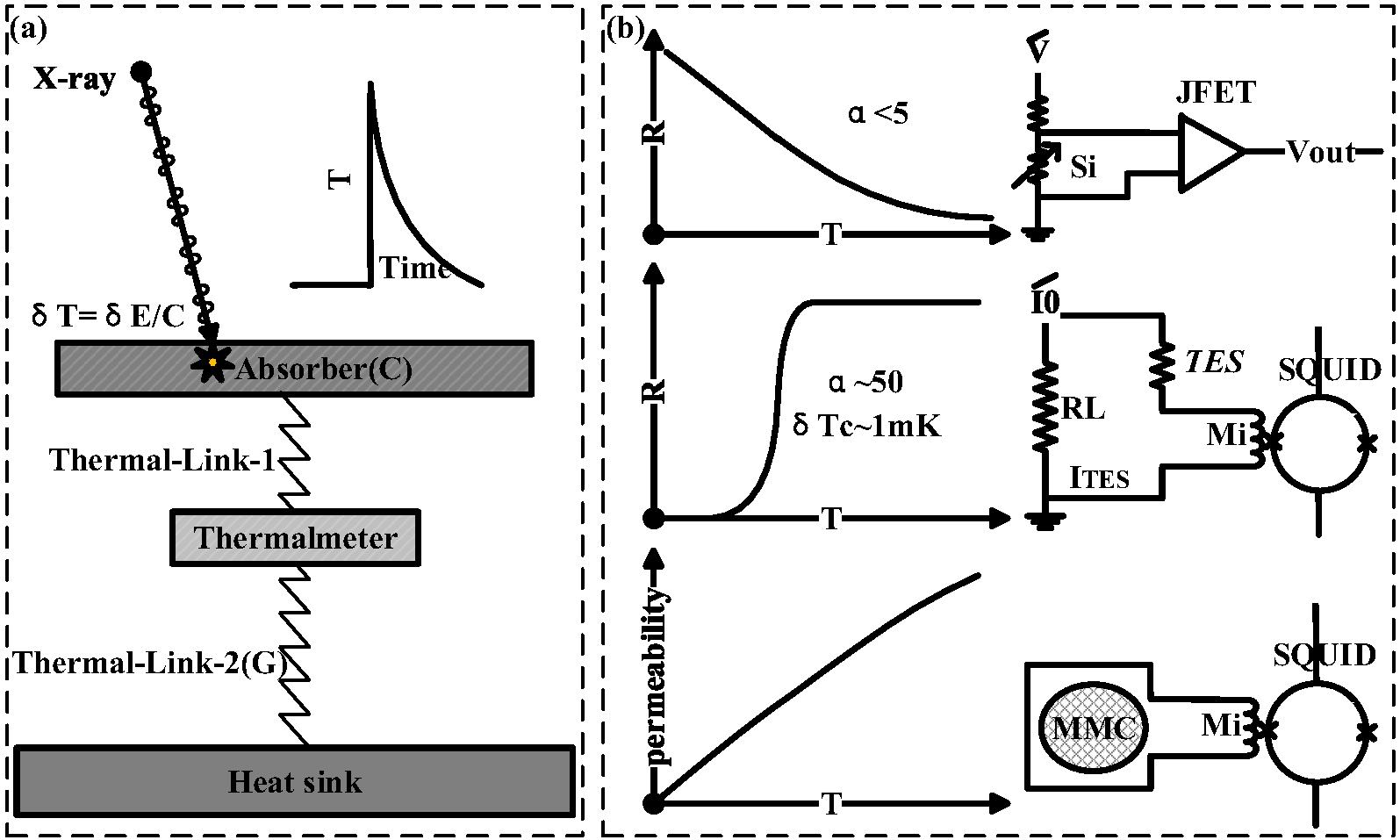}
\caption{Principle of microcalorimeters: Figure (a) shows the main structure and working principle of microcalorimeters. It consists of four parts: absorber, thermometer, thermal weak link and heat sink. X-ray are absorbed and converted into thermal signals, and then converted into an electrical signal. The thermal signal will finally be conducted to the heat sink, and then return to the equilibrium state to wait for the next incident particle. Figure (b) shows three thermometer that can be applied to microcalorimeters. According to the impedance and signal type of each thermometer, the type of signal amplifier is also different.}
\label{fig1}
\end{figure*}
\section{Introduction}\label{sec.I}

TES is an extremely sensitive temperature sensor, The main structure of TES is a superconducting thin film. Under certain temperature and bias current, it can work in the transition state between superconducting and non-superconducting\cite{1-TES-2005}. Since the transition region is very narrow, the resistance of TES is extremely sensitive to the change of temperature. The temperature coefficient of resistance($\alpha_I$) of TES is much higher than that of traditional semiconductor thermometers, so it has been used in many cutting-edge scientific researches\cite{2-TESreview-2015,3-TESSRON-2021,4-Synchrotron-2006,5-Synchrotron-2015,6-Synchrotron-2017,7-Ledge-2021,8-JPTES-2020}. For the sake of brevity, this article will mainly introduce the work in the field of X-ray and $\gamma$ ray.

In the X-ray energy range, the application fields of TES mainly include synchrotron-based beamline endstations\cite{4-Synchrotron-2006,5-Synchrotron-2015,6-Synchrotron-2017,9-TRXAS-2013,10-TRXES-2015,11-TRXES-2017,12-TRXES-2016}, accelerators, EBIT\cite{13-Hadron-2016,8-JPTES-2020,14-kaonic-2020,15-EBIT-2017}, X-ray astronomy\cite{16-MX-2020,17-XIUF-2016,18-HUBS-2020}and electron microscopy\cite{19-SEM-2020}, \textit{etc}. The X-ray flux of the beamline stations based on synchrotron radiation and free electron laser is much higher than that of laboratory-level sources. In addition to the rapid analysis of general materials, there is an urgent demand for energy spectrum measurement of dim and non-point X-ray sources, so the demand for TES X-ray detectors is increasing\cite{4-Synchrotron-2006,5-Synchrotron-2015,6-Synchrotron-2017,8-JPTES-2020}. TES X-ray detectors have been used in research areas such as low-Z element X-ray emission spectroscopy (XES)\cite{5-Synchrotron-2015}, X-ray absorption near-edge spectroscopy \cite{25-NEXAS-2019,26-NEXAS-2019,27-NEXAS-2017}, time-resolved X-ray absorption and emission spectroscopy\cite{10-TRXES-2015,11-TRXES-2017,12-TRXES-2016}, resonant soft X-ray scattering (RSXS)\cite{28-RSXS-2020}. On particle accelerators, TES X-ray detectors have been used for meson atomic energy spectrum measurements such as $\pi$ and $\kappa$\cite{8-JPTES-2020}. On high charge state ion traps, TES X-ray detectors have been and will be used for high charge state atomic spectroscopy measurements\cite{15-EBIT-2017,29-EBIT-2014,30-EBIT-2009,31-EBIT-2019}. In X-ray astronomy, TES X-ray detectors are used in Micro-X sounding rocket experiments, and will be used in space science observation platforms such as ATHENA and HUBS satellites\cite{16-MX-2020,17-XIUF-2016,18-HUBS-2020}. In the applications related with electron microscopy, TES X-ray detectors are introduced into scanning electron microscopy for elemental distribution and valence analysis with high spatial resolution\cite{19-SEM-2020,33-DR-2014,34-SEM-2021}.

In the $\gamma$-ray energy range, the application fields of TES mainly include nuclear clocks\cite{20-NuCl-2019}, nuclear security inspection\cite{21-securaty-2009,22-256gamma-2015} and nuclear medicine. The nuclear clock uses the nuclear energy level change to determine the period. Compared with the traditional atomic clock, the energy level of nucleus is less affected by the environment. Currently, the main selected nuclide is $\ce{^{229}Th}$. In order to determine the energy level, a number of institutions in Japan used TES $\gamma$-ray detectors to accurately measure the relevant energy levels\cite{20-NuCl-2019}. TES $\gamma$-ray detectors can also be used for national security inspections, the most important signal of Highly Enriched Uranium(HEU) is the 185.7 keV $\gamma$-ray, which is almost identical to the 186.1 keV $\gamma$-ray emitted by $\ce{^{226}Ra}$ present in other materials, the two spectral lines can be easily identified using the TES $\gamma$-ray detector\cite{21-securaty-2009}. In addition, NIST used their TES $\gamma$-ray detector to accurately measure the spectral lines of the plutonium isotope mixture\cite{35-securaty-2015}. In the field of nuclear medicine, TES $\gamma$-ray detectors can be used for the detection of $\gamma$-ray from radioactive materials.

After years of development, ShanghaiTech University and its cooperative institutions have developed a TES X-ray detector\cite{36-Physica-2021}. Using this detector, elemental analysis is performed on the energy spectrum of a variety of materials. This paper will show energy spectra of coin, stone and high-purity niobium. The TES detector for soft X-ray has an absorber with a thickness of several microns, with a relatively inadequate ability to trap photons above 10keV. In order to expand the energy range, the research team combined submillimeter-sized lead-tin alloy balls with TES to perform energy spectrum measurement for $\gamma$-ray, and has obtained an energy resolution of 161.5 eV at 59.5 keV.

\section{Theoretical and instrument descriptions}\label{sec.II}
The TES X/$\gamma$-ray detector is a type of microcalorimeter. In this section, the principle and types of the microcalorimeter are introduced first, and then come to the structure of the TES X/$\gamma$-ray detector and its related systems are briefly introduced. Then, several energy spectrums are shown, and finally the structure of the TES $\gamma$-ray detector developed upon this basis is introduced.

\subsection{Principle of Microcalorimeter}\label{A}
Microcalorimeter is a new type of detector based on thermal signal proposed by Moseley, Mather and Dan McCammon in 1984\cite{37-microcalorimeter-1984}. As is shown in left part of Figure \ref{fig1}, the microcalorimeter consists of four parts: absorber, thermometer, thermal weak link and heat sink. When the incident particle is stopped by the absorber, its kinetic energy be converted to thermal energy, causing a temperature rise of the absorber: $\delta{T}\propto\delta{E}/C$, where $C$ is the heat capacity. By using a thermometer $\delta{R}\propto\delta{T}$, the energy of incident particle $\delta{E}\propto\delta{R}*C$ can be inferred. Generally, the linear region of the detector is selected for energy spectrum measurement, so $\delta{E}=k*\delta{R}*C$. After calibration with X-ray source, the slope $k$ can be obtained. Unlike the non-equilibrium detectors, the energy resolution of TES $\Delta{E_{\rm FWHM}}$ is independent of the X-ray energy $E$: $\delta{E_{\rm FWHM}\propto\sqrt{4{k_{ \rm B}}{{T_0}^2}C/{\alpha_I}}}$, where $\alpha_I$ is the temperature coefficient of resistance\cite{38-MicroCal-2005}. According to this equation, in order to obtain higher energy resolution, a lower temperature, smaller heat capacity and higher $\alpha_I$ are required. Combining the two  factors such as the cooling power of the cryogenics system and the performance of the energy spectrometer, the microcalorimeter generally works at 100 mK or lower temperature. For heat capacity, the absorber volume should be as small as possible under the premise of ensuring sufficient quantum efficiency and detection area.

According to the types of thermometers, microcalorimeters are mainly divided into three types: semiconductor microcalorimeters\cite{39-SemMC-2005}, TES microcalorimeters\cite{1-TES-2005} and Metallic Magnetic Calorimeters (MMC)\cite{40-MMC-2005}. Semiconductor microcalorimeters use boron-doped silicon or transmuted germanium as temperature sensors, with a low sensitivity of the change of temperature. Because of its large impedance, JFET can be used as the signal amplifier. TES have a very high temperature sensitivity. However, the transition edge is very narrow, so it is prone to be saturation, and have a relatively poor linearity. Because of its low impedance, a Superconducting QUantum Interference Device(SQUID) is required as a signal amplifier. MMC uses paramagnetic materials as temperature sensors, it have no saturation problems and have a very good linearity. However, due to the large heat capacity, it require lower operating temperature. Besides, it is difficult to multiplex the readout circuit, so MMC is still under development.

Compared with semiconductor thermometers, the improvement of the temperature sensitivity 10 to 100 times by TES makes it more flexible for structural design and material selection. For example, a high specific heat material such as gold(Au) can be selected as the absorber. This change greatly improves its count rate, also greatly expanding the range of material selection, allowing the use of microfabrication to obtain large arrays. The reduction of the heat capacity limitation of the absorber also makes it possible to bury the source in the absorber, so several research institutes use this feature to conduct neutrino mass measurement research\cite{41-neotrino-2016}.

Selection of the material of the absorber is one of the main points of this work. The time constant of microcalorimeter can be express as $\tau=C/G$, where $G$ is thermal conductance. Once the structure and the working temperature are determined, the value of $G$ is relatively fixed. According to the formula, a smaller heat capacity corresponds to a smaller time constant. According to the expression of $\delta{E_{\rm FWHM}}$, the smaller the heat capacity of the absorber, the higher the energy resolution. A small heat capacity therefore corresponds to a high time resolution and a high energy resolution. The heat capacity of the absorber is $C=c_v\times{S}\times{H}$, where $c_v$ is the specific heat capacity per unit volume, $S$ is the area, and $H$ is the thickness. For X-ray and $\gamma$-ray, the higher the energy, the stronger the transmission ability. After the upper limit of the expected energy $E_{max}$ is determined, the thickness $H_{min}$ corresponding to $E_{max}$ and the predetermined quantum efficiency (such as 70\%) can be calculated, and then the maximum photosensitive area $S_{max}$ that can be obtained. In order to obtain a larger photosensitive area, the absorber shall be chosen to have the best trade off among the high Z and low specific heat capacity. The specific heat capacity at extremely low temperature is mainly contributed by electrons and phonons, materials with higher specific heat capacity also have better conductivity. Therefore, from the perspective of maximizing the photosensitive area, it is desirable to use materials with lower electric conductivity, for example, semi-metals such as bismuth or superconducting materials(such as lead). To improve the conductance, a layer of gold or copper is added to the absorber to improve thermal conductivity.

\subsection{Structure of TES X/$\gamma$ ray detector}\label{B}
The TES X/$\gamma$  ray detector includes TES chip, input-SQUID and low temperature block. The TES chip converts X-ray into current signals, and the input-SQUID converts the current signals  into voltage signals. The functions of low temperature block mainly include low temperature circuit connection, magnetic shielding, electromagnetic shielding, heat sink and infrared window.

\begin{figure}[!htb]
\includegraphics[width=.96\hsize]{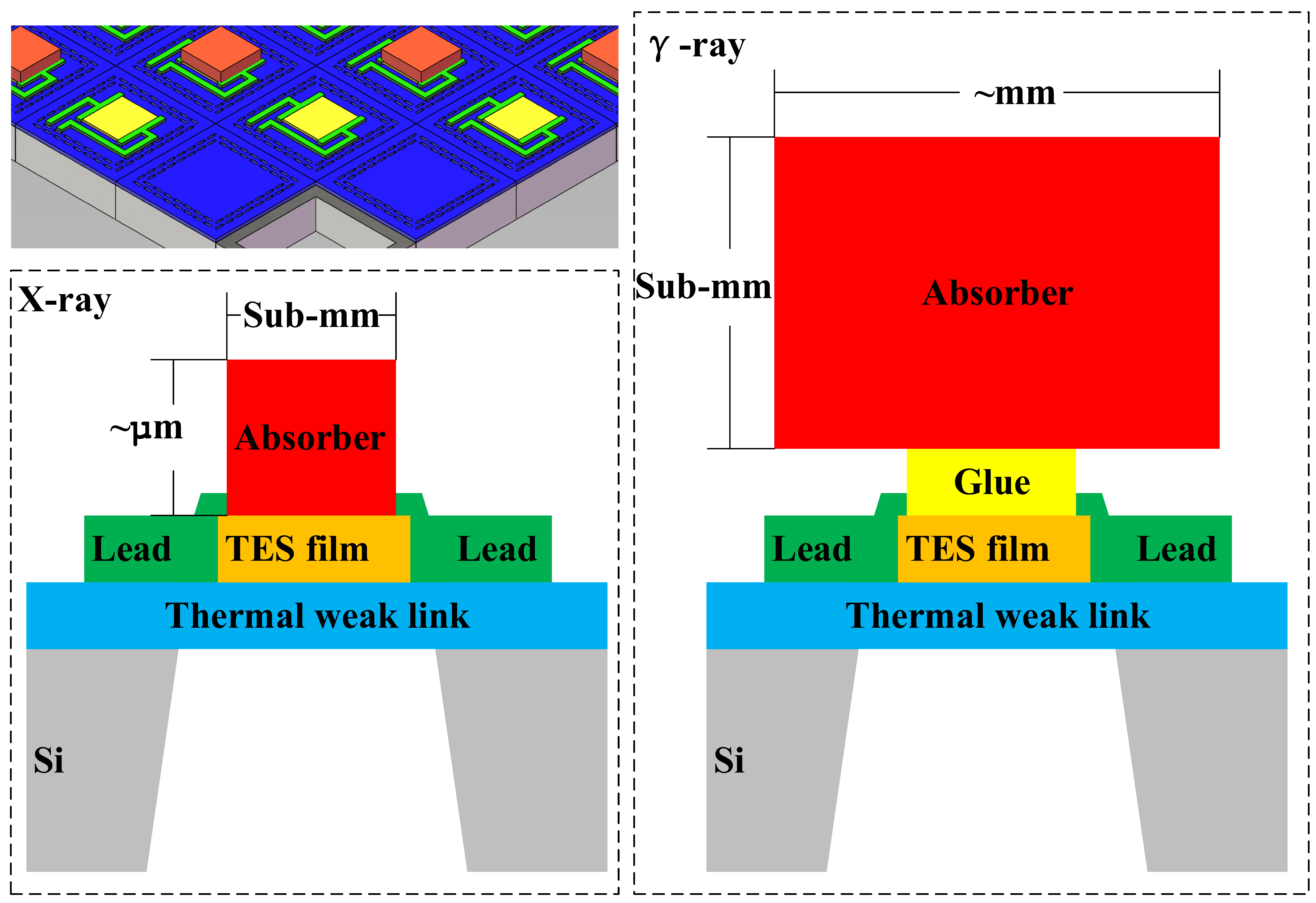}
\caption{The typical structure of a TES chip, the silicon substrate is used as the support structure and heat sink, the suspended silicon nitride is used as thermally weak link structure, an superconducting film is used as a thermometer, high Z-value metal as absorber.}
\label{fig2}
\end{figure}

The typical structure of a TES sensor chip is shown in the Figure \ref{fig2}: the silicon substrate is used as the support structure and heat sink, and the thickness of the silicon material is generally between 300 microns and 500 microns. The suspended silicon nitride is used as a thermally weak link structure, which can be realized by removing all the backside silicon by deep silicon etching. The thickness of the silicon nitride is in the order of submicron. The molybdenum-copper bilayer superconducting material is used as a thermometer, and the total thickness is within the order of submicron. Superconducting materials such as molybdenum or niobium are used as wires with a typical thickness in the order of 100 nanometers. Using high-Z value materials as the absorber, the thickness in the X ray range is typically of the order of few micrometers, while for $\gamma$-ray, it shall reach the submillimeter level.

\begin{figure}[!htb]
\includegraphics[width=.96\hsize]{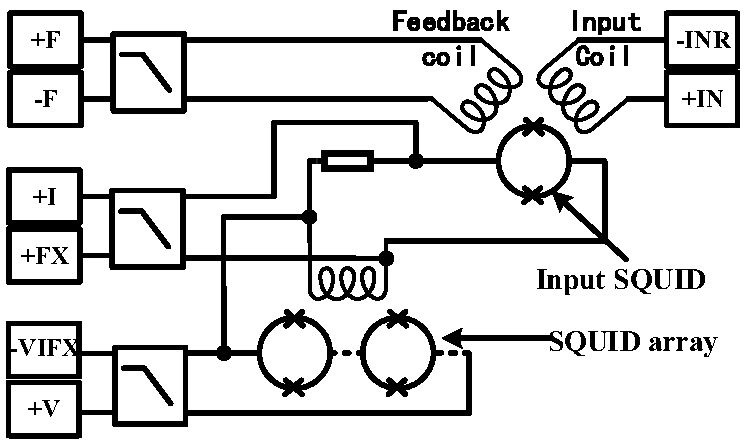}
\caption{A structure diagram of the none-multiplexed SQUID. The input-SQUID turn the current signal into a voltage signal, and the SQUID array on the lower left amplifies the signal further to increase signals to noise ratio\cite{42-SQUID-201}.}
\label{fig3}
\end{figure}

The normal resistance of the TES chip used for X/$\gamma$  ray detection is low, generally at the level of 10 m$\Omega$, so it use input-SQUID chip to convert the current signal into voltage signal. A typical non-multiplexed SQUID structure is shown in Figure \ref{fig3}. A single-stage SQUID consists of two Josephson junctions with the same structure. When the current in the TES changes, the voltage across the device changes, and the voltage signal is transmitted to the low-temperature signal amplifier through the cable for further amplification and feedback.

The low temperature block is responsible for electrically connecting the TES chip and the input-SQUID, and is also responsible for the thermal connection and the refrigerator (cold head), the shielding of electromagnetic/magnetic fields, etc. Low temperature block is limited by chip structure and scientific applications, and its structure is various.

\subsection{Auxiliary system of TES X/$\gamma$ ray detector}\label{C}
The detector needs a cryogenics system, a low-temperature signal amplifier, and a data acquisition and analysis system(DAQ) to realize the energy spectrum measurement. Here is a brief introduction for them.

\begin{figure*}[!htb]
\includegraphics[width=.7\hsize]{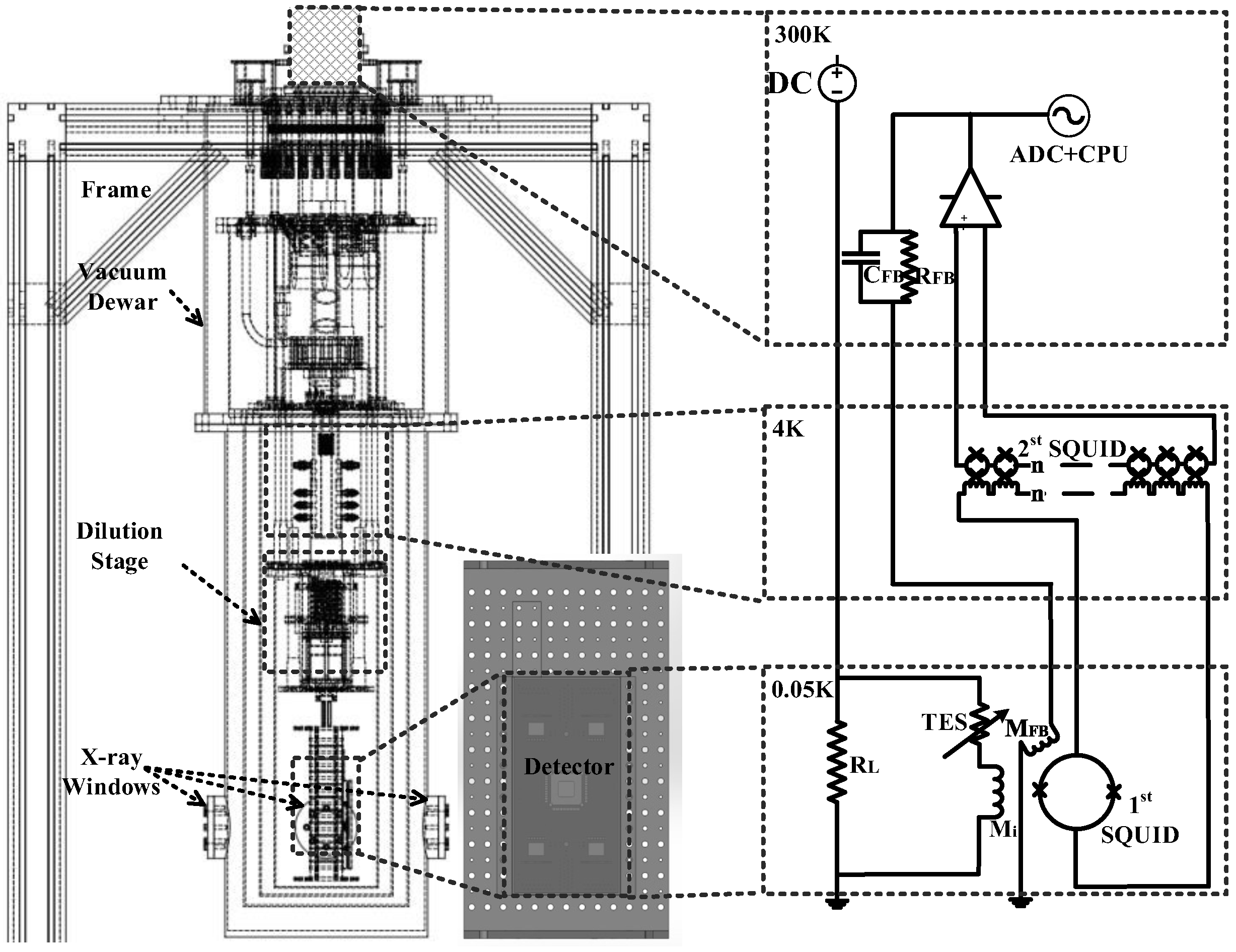}
\caption{Structure of the auxiliary system: The main structure of the auxiliary system is a set of dilution refrigerator. The Low temperature block is installed on the 50 mK cold plate. X-ray are irradiated to the TES X-ray detector through the X-ray window. The signal is amplified and transmitted from 4 K to 300 K by the low-temperature signal amplifier, and then reaches the DAQ.}
\label{fig4}
\end{figure*}

\textbf{Cryogenics system:}
The refrigeration system provides with low temperature, low vibration, low magnetic field, low electromagnetic interference environments for the TES X-ray detector, mainly including thermostat, temperature control system, vibration isolation structure, magnetic/electromagnetic field shielding system. As is mentioned above, the thermostat needs to reach a temperature of 100 mK or lower. In order to reduce the temperature fluctuation of the low-temperature X-ray detector, it is generally necessary to control the temperature fluctuation of the thermostat to the order of $\mu$K.

\textbf{Low-temperature signal amplifier:}
Signal amplification systems based on non-multiplex SQUID, it generally use a SQUID array at the 4 K cold plate to further amplify the signal. Its main function is to suppress the noise of the operational amplifier in the room temperature circuit. Figure \ref{fig4} shows the feedback amplification principle. The current signal $\delta{I_{\rm{input}}}$ first generates a magnetic field signal in the SQUID loop, and then forms a voltage signal. The voltage signal is further amplified by the 4K SQUID array and operational amplifier at the room temperature. The operational amplifier passes the signal through the resistor $R_{\rm{f}}$ to form a negative feedback, and the feedback coil forms a reverse changing magnetic field in the SQUID loop, and finally reaches a state where the feedback current $\delta{I_{\rm{f}}}$ and $\delta{I_{\rm{input}}}$ is proportional to each other: $\delta{I_{\rm{f}}}=k_{\rm{f}}\times\delta{I_{\rm{input}}}$. And the room temperature terminal voltage signal is proportional to the low temperature current signal $\delta{V_{ \rm{300K}}}=\delta{I_{\rm{f}}}\times{R_{\rm{f}}}=k_{\rm{f}}\times\delta{I_{\rm{input}}}\times{R_{\rm{f}}}$.

\textbf{DAQ:}
The data acquisition and analysis system mainly includes an analog-to-digital conversion module (ADC), a waveform reconstruction module and an energy spectrum analysis module.

\begin{figure}[!htb]
\includegraphics[width=.96\hsize]{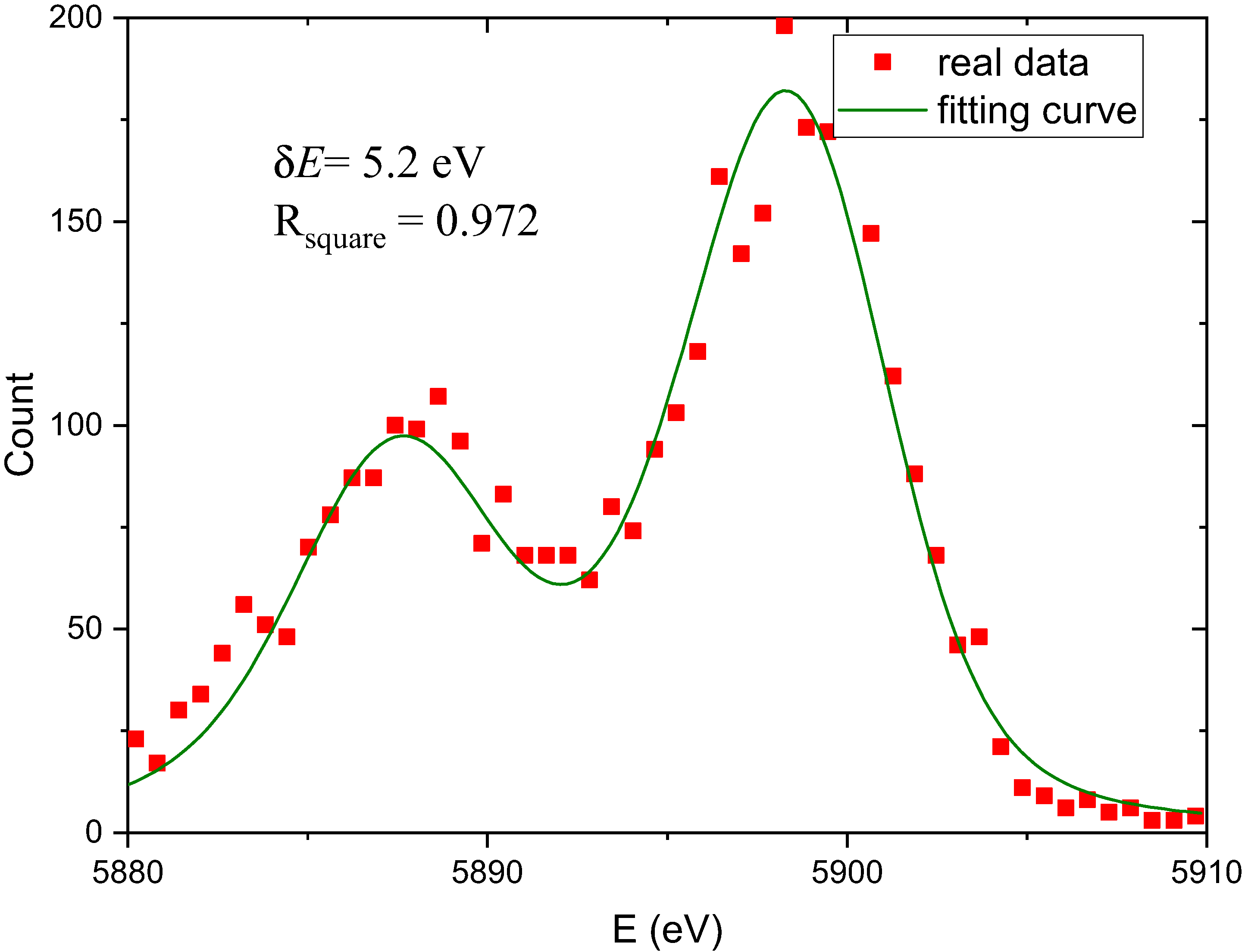}
\caption{The X-ray energy spectrum of $\ce{^{55}Fe}$ radioactive source. $\ce{^{55}Fe}$  emits two X-ray spectral lines with energies of 5900 eV and 5890 eV, and the detector can clearly distinguish them. After fitting, the detector obtained an energy resolution of about 5.2 eV around 5900 eV with an $\rm{R_{square}}=0.972$ .}
\label{fig5}
\end{figure}

\begin{figure*}[!htb]
\includegraphics[width=.7\hsize]{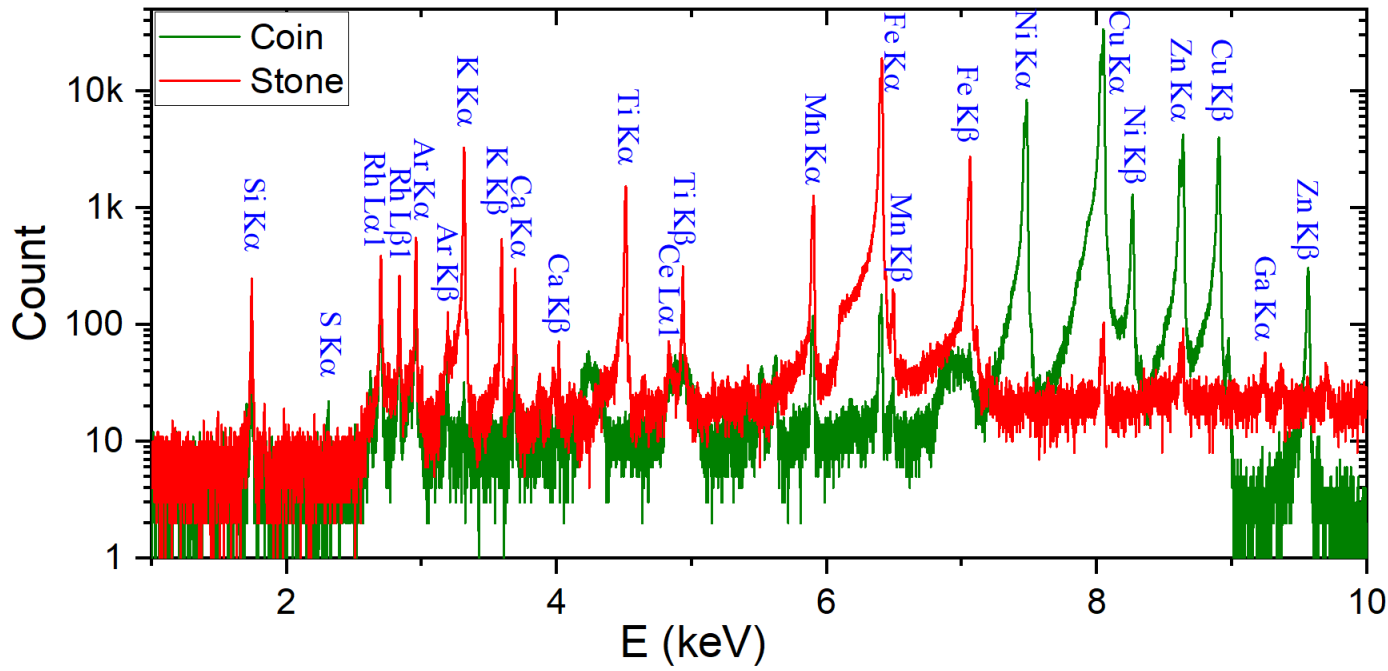}
\caption{The energy spectrum obtained by irradiating complex composition materials such as coins and stones with a Rh target. The $K_{\alpha}$ and $K_{\beta}$ spectral lines of copper, zinc, nickel, iron, manganese, calcium, titanium, potassium, silicon, argon and other elements can be clearly seen from the energy spectrum. Due to the difference in composition, coins and stones have very obvious differences in the relative intensities of the spectral lines.}
\label{fig6}
\end{figure*}

\begin{figure*}[!htb]
\includegraphics[width=.7\hsize]{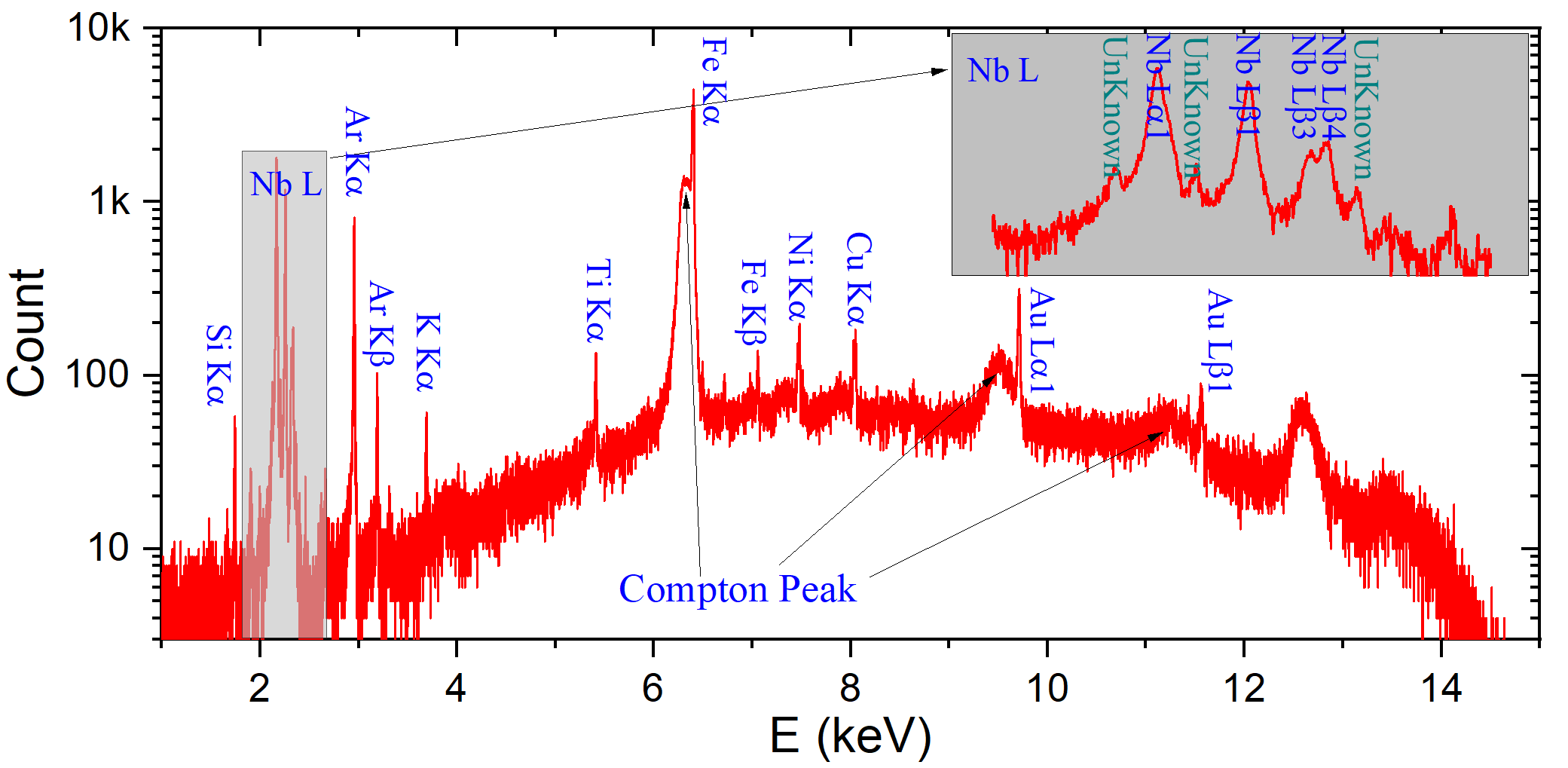}
\caption{The energy spectrum obtained by irradiating high-purity niobium with an Au target X-ray tube, from the energy spectrum that $L_{\alpha 1}$, $L_{\beta 1}$, $L_{\beta 3}$, $L_{\beta 4}$ of Nb element lines can be clearly seen. On the energy spectrum, we can see the obvious $L_{\alpha 1}$ and $L_{\beta 1}$ spectral lines of gold, and the $K_{\alpha}$ line of iron element can be seen at the same time.}
\label{fig7}
\end{figure*}

\subsection{TES X-ray detectors at Shanghaitech University and measured spectrum}\label{D}

Several sets of synchrotron radiation facilities and free electron laser facilities have been constructed or are under construction around Shanghaitech University, therefore, many beamline end stations are also being constructed simultaneously. Based on the energy spectrum measurement requirements of several end stations, Shanghaitech University has built a set of TES X-ray detectors for the X-ray band to meet the needs. The detector uses a $\rm{4\times4}$ pixel TES sensor chip, and the area of a single pixel is about $\rm{0.5mm\times0.5mm}$ . The TES thin film is a molybdenum-copper bilayer film with a transition temperature of about 70 mK, with the molybdenum leads. The absorber layer is 2.4 micron thick bismuth with a detection efficiency of 70\% at 5.9 keV. The input-SQUID chip used a single-stage SQUID. The the refrigeration system has a minimum working temperature of 23 mK. The low-temperature amplifier adopts 16 SQUID arrays connected in series, and the room temperature feedback amplifier circuit have a bandwidth of 1 MHz. An ADC board based on the PXIe chassis is used in DAQ. After the signal is transmitted to the computer, the pulse height is fitted by offline software, Then the heights are calculated and calibrated, and the X-ray energy spectrum is finally obtained. After preliminary tests, the detector has an energy resolution of 5.2 eV at 5.9 keV. The energy spectrum and fitting curve are shown in Figure \ref{fig5}.

Two different cases is shown here. The first case is that the composition of the sample is extremely complex, and the relative intensities of the spectral lines of many elements need to be determined in a wide energy range. The second case is the measurement of impurity elements in high-purity samples, which needs to find faint impurity spectral lines in a strong scattering background environment. In the first case we used coins and stones as samples, in the second one we take samples of high-purity (99.95\%) niobium.

As is shown in Figure \ref{fig6}, X-ray from Rh target tube irradiates on coins and stones, and the fluorescence spectrum is collected by the TES X-ray detector. The Rh target voltage is set to 15 kV. Since the K absorption edge energy level of Rh is about 23 keV, the output X-ray is mainly generated by the bremsstrahlung process. The energy spectrum is relatively smooth, and the photon flux changes slowly with the energy.

The main content of the coin used for testing is brass, with a nickel-plated pattern in some areas. From the X-ray fluorescence spectrum, we can see very obvious $K_{\alpha}$ and $K_{\beta}$ spectral lines of copper, zinc and nickel, and at the same time, lower content of iron and much lower content of manganese and silicon can be seen. On the spectrum, we can see the obvious $L_{\alpha1}$ and $L_{\beta1}$ spectral lines of Rh, which come from the reflection of the input X-ray by the coin. There are obvious Ar $K_{\alpha}$ and $K_{\beta}$ spectral lines in the energy spectrum, which come from argon in air.

The stone used for the test is red in color and has a high iron content. From the X-ray fluorescence spectrum, we can see very obvious $K_{\alpha}$ and $K_{\beta}$ spectral lines of iron, manganese, calcium, titanium, potassium, and silicon elements. And at the same time, lower content iron and lower levels of Ce , copper, zinc and gallium can be seen. On the spectrum, obvious Rh $L_{\alpha1}$ and $L_{\beta1}$ spectral lines can be seen, which come from the reflection of the input X-ray by the stone. There are also obvious Ar $K_{\alpha}$ and $K_{\beta}$ spectral lines.

The Au target X-ray tube was irradiated to high pure niobium. The Au target voltage is set to 15 kV. Since the L absorption edge energy level of Au is lower than 15 keV, the X-ray output of the X-ray tube is mainly composed of a smooth energy spectrum generated by bremsstrahlung and a sharp characteristic peak.

From the X-ray fluorescence spectrum, the L characteristic line group of niobium can be seen very clearly. From the enlarged subgraph of Fig.\ref{fig7}, $L_{\alpha 1}$, $L_{\beta 1}$ $L_{\beta 3}$, $L_{\beta 4}$ and other spectral lines of niobium element can be seen, and there are spectral lines around them that cannot be corresponded in the database. The $K_{\alpha}$ and $K_{\beta}$ lines of Ar can also be seen on the X-ray fluorescence spectrum. The $K_{\alpha}$ lines of low content silicon, potassium, titanium, nickel, and copper shown in the energy spectrum indicate that they are the main components of impurities in the metal niobium. On the energy spectrum, we can see obvious $L_{\alpha 1}$ and $L_{\beta 1}$ spectral lines of gold, and $K_{\alpha}$ lines of iron element, they come from the reflection of niobium for incoming X-ray. The ratio of $K_{\alpha}$ and $K_{\beta}$ spectral lines of iron are seriously deviated from the theoretical values, the detail reasons need to be further explored in future work.

On the energy spectrum of niobium, at the low-energy end of the $L_{\alpha 1}$, $L_{\beta 1}$ spectral lines of gold and  $K_{\alpha}$ of iron, bulge structures can be seen. These structures are due to the Compton scattering when the X-ray irradiate the sample,and eventually form a bulge structure at the low-energy end of the elastic scattering peak. Since the Compton scattering bulge is always accompanied by the elastic scattering peak, its existence can be used to judge whether it is corresponding to the elastic scattering peak or the characteristic spectral line. For example, at the low energy end of the spectral line of $L_{\alpha 1}$ of niobium element, no bulge structure can be seen, so it can be judged that $L_{\alpha 1}$ of niobium comes from niobium in the sample. And at the low-energy end of the iron $K_{\alpha}$ spectral line, an obvious bulge structure can be seen, so it can be judged that most of the iron element $K_{\alpha}$ spectral line does not come from the sample, but from the iron-containing vacuum parts around the sample.

\begin{figure*}[!htb]
\includegraphics[width=.7\hsize]{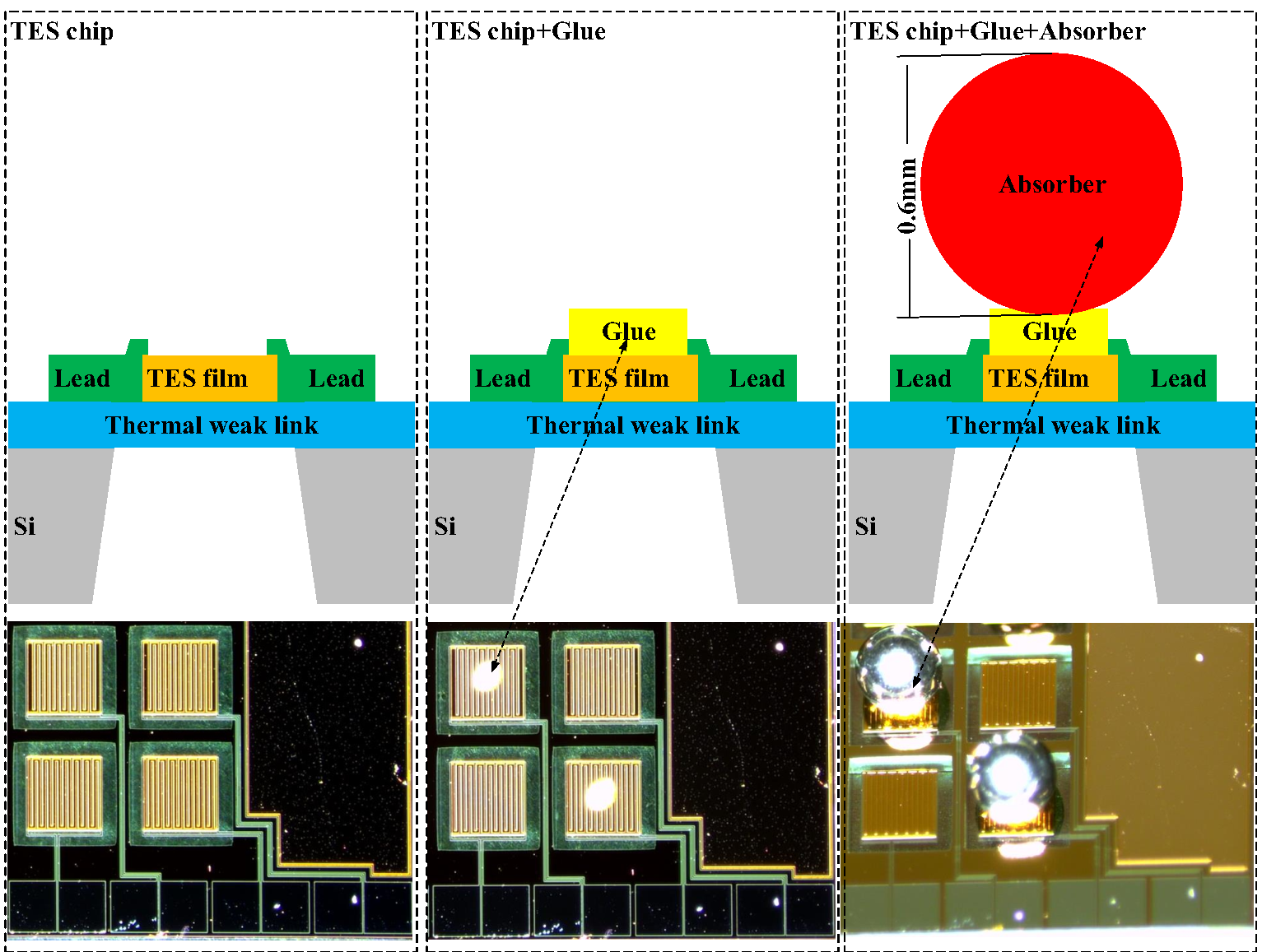}
\caption{The fabrication process of the TES $\gamma$-ray sensor: firstly prepare a TES sensor array, next put a very small amount of epoxy resin on the TES sensor, then the pre-processed lead-tin alloy balls are bonded by the robotic arm.}
\label{fig8}
\end{figure*}

\subsection{Fabrication of $\gamma$-ray detector based on TES}\label{E}
In order to extend the measurement energy range to 100 keV or even 200 keV, the thickness of the absorber needs to be increased to sub-millimeter level. The lead-tin alloy(63\% Pb, 37\% Sn) is used as the absorbtion material. The first reason for choosing this material is that it is a superconducting material with a small specific heat capacity at low temperature\cite{2-TESreview-2015}. At the same time, lead and tin have a strong absorption effect on X/$\gamma$-ray. The thermal transport process of incident X-ray on different positions of the absorber is not the same, which will cause X-ray with the same energy to generate different pulse shapes. In order to reduce this effect, the shape of the absorber is processed into a spherical shape. The spherical structure also reduces the bonding point area, which can reduce the difficulty of the bonding process. The fabrication process of the entire TES $\gamma$-ray sensor is shown in Figure \ref{fig8}: firstly prepare a TES sensor array, then put a very small amount of epoxy resin on the TES sensor, afterwards, the pre-processed lead-tin alloy balls are bonded by the robotic arm. After the epoxy resin is cured, replace the TES X-ray sensor mentioned in the previous chapter with this TES $\gamma$-ray sensor. Finally a set of TES $\gamma$-ray detectors for $\gamma$-ray energy range is obtained.

\begin{figure*}[!htb]
\includegraphics[width=.7\hsize]{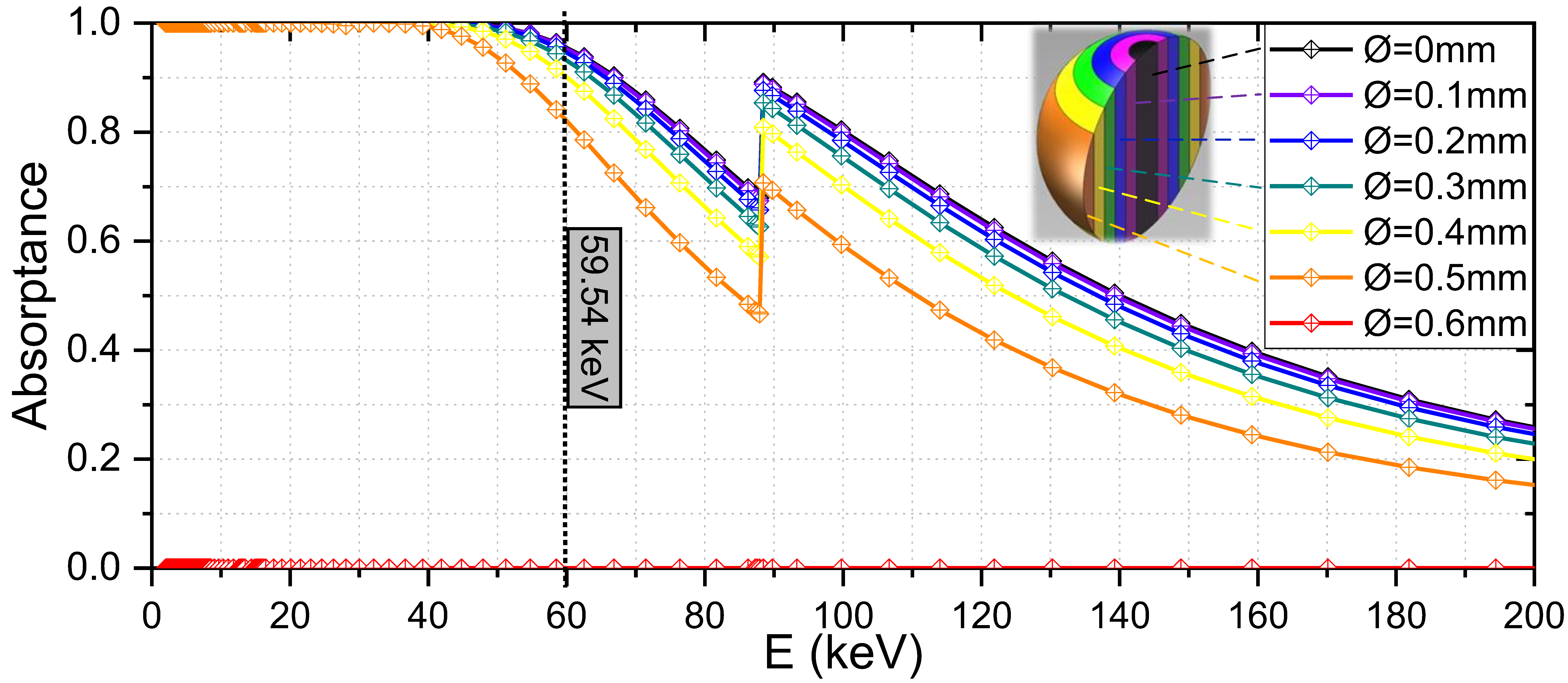}
\caption{Absorption efficiency at different distances from the ball axis, the diameter of the ball is 0.6 mm, and the material is lead-tin alloy.}
\label{fig9}
\end{figure*}

For the size and shape of the absorber, the absorbability of lead-tin alloys with different thicknesses to X-ray and $\gamma$-ray with different energies was calculated through the NIST database(PhysRefData). After preliminary calculations, a sphere with a diameter of 0.6 mm can obtain high detection efficiency in most positions, so the results presented in this paper are based on a lead-tin alloy absorber with a diameter of 0.6 mm. Figure \ref{fig9} shows the absorption efficiency curves of different distances from the ball axis. It can be seen that the position within 0.5 mm from the axis still has a high blocking ability, and the blocking ability is sharply decline in the area beyond 0.5 mm.

\section{Results and discussion}\label{sec.III}
\subsection{Brief description of experimental setup}\label{A}
We use a $\ce{^{241}Am}$ as the $\gamma$-ray source. Coating $\ce{^{241}Am}$ on a silver-plated metal sheet, and then wrapping $\ce{^{241}Am}$ with aluminum foil. It can effectively reduce the risk of $\ce{^{241}Am}$ pollution to the surrounding environment, and at the same time can reduce the signal intensity of $\alpha$ particles and $\beta$ particles, the thickness of the aluminum foil is about 0.1 mm. It is placed inside the refrigerator 5 millimeters away from the TES $\gamma$-ray detector, and the surrounding is covered with lead, and then the dilution refrigerator is cooled down. After the temperature reached about 23 mK, the temperature resistance(R-T) curve of the TES device was measured. The measured transition temperature was about 70 mK. Biasing TES on the transition edge, finally, the energy spectrum of the X-ray and $\gamma$-ray generated by the $\ce{^{241}Am}$ source is measured.

\subsection{Time constant and count rate}\label{B}
\begin{figure*}[!htb]
\includegraphics[width=.7\hsize]{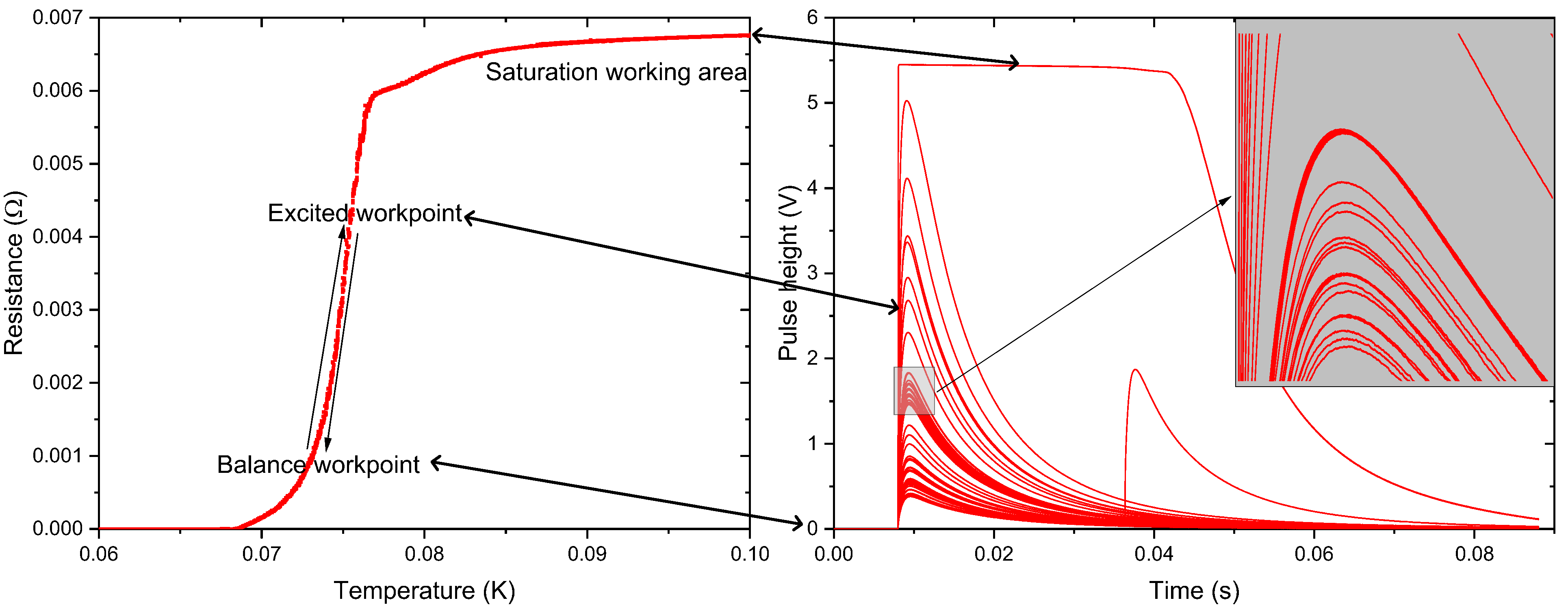}
\caption{Left plot: The R-T curve of the TES $\gamma$-ray detector. When collecting the energy spectrum, it is biased at the position of 1 m$\Omega$. $\gamma$-ray will cause the temperature and resistance of the detector rise. Right plot: The pulse signals generated by different energy X-ray and $\gamma$-ray have a rising edge width of about 0.2 ms and a falling edge width of about 80 ms. It can be seen on the zoomed-in image that, corresponding to the nuclear spectral line from $\ce{^{241}Am}$  with an energy of 59.5 keV, the signal is most dense around 1.845 V.}
\label{fig10}
\end{figure*}

The response of the TES $\gamma$-ray detector to a single $\gamma$-ray is shown in Figure \ref{fig10}. The width of the rising edge is about 0.2 ms, and the width of the falling edge is about 80 ms. The falling edge of the pulse signal decays quasi-exponentially. By fitting, the average time decay constant is $\tau = 8.54\pm0.06$ ms. After 10 $\tau$, the signals return to the baseline level. According to the estimation of this time constant, the maximum count rate of a single pixel of the detector is around 10 CPS. This means that the detector is not suitable for high count rate measurement scenarios.

\subsection{Linear Intervals and Saturation Energy}\label{C}
\begin{figure*}[!htb]
\includegraphics[width=.7\hsize]{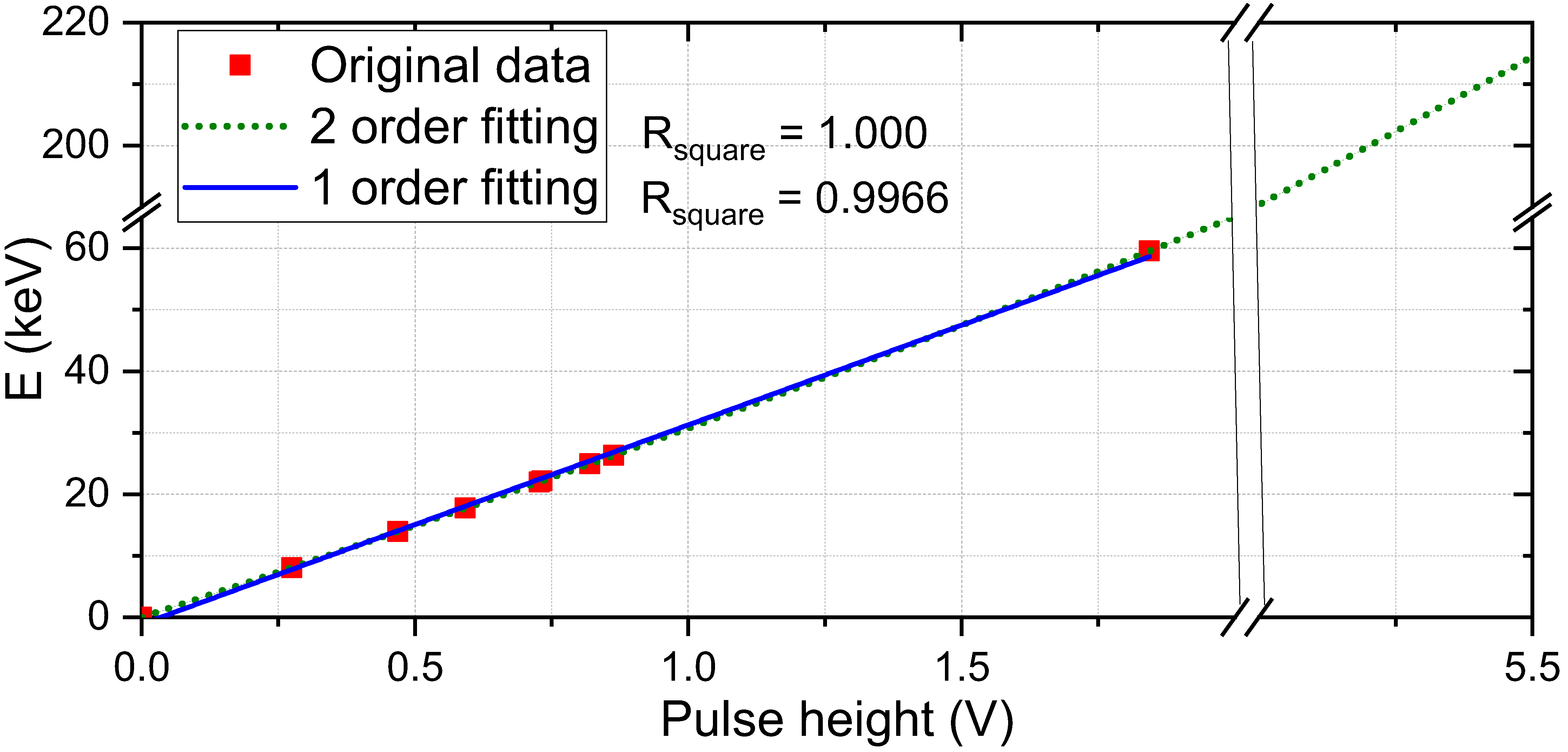}
\caption{There is a certain deviation in the fitting of the quadratic function, the fitting effect of the quadratic function is excellent, and the $\rm{R_{square}}$ reaches 1.000.}
\label{fig11}
\end{figure*}

$\ce{^{241}Am}$ decays to Np after emitting $\alpha$ particles with energy of about 5 MeV. And a $\gamma$-ray of 59.5 keV with a 35.92\% probability. When Np is excited, it will emit a certain amount of fluorescence lines corresponding to the $K$ energy shell and the $L$ energy shell. Since there are supporting structures containing elements such as silver, copper, iron, and chromium around the source, the corresponding fluorescence lines can also be seen in the energy spectrum. The energy of each spectral line can be obtained by querying the X-ray fluorescence database and the manual of $\ce{^{241}Am}$. According to the settings of the experimental device, it is predicted that the number of pulses corresponding to 59.5 keV photons is the largest. It can be judged from the right image of Figure \ref{fig10} that the pulse signal with a height of about 1.845 V corresponds to this energy. Through linear estimation, the energy spectrum is found with $K_{\alpha}$ of copper, $L_{\alpha 1}$\&$L_{\beta 1}$ of Np, $K_{\alpha 1}$\&$K_{\alpha 2}$\&$K_{\beta}$ of silver and 26.3 keV\&59.5 keV nuclear transition lines of $\ce{^{241}Am}$. Useing the pulses height corresponding to these spectral lines to fit the energy values in the database, as shown in the Figure \ref{fig11}, there is a certain deviation in the fitting of the one power function($\rm{R_{square}=0.9966}$), the fitting effect of the quadratic function is excellent($\rm{R_{square}=1.000}$), so the relationship between the pulse height and the energy can be expressed by a quadratic function,

It can be seen from the right plot of Figure \ref{fig10} that the falling edge of the pulse whose height is lower than 5.5 V basically decays exponentially. When the incident particle energy is too high, the pulse will appear saturated with a flat top of about 5.5 V. After a period of time, it will evolve to the equilibrium state at an exponential decay rate. It can be seen from the Figure \ref{fig11} that the pulse height of 5.5 V corresponds to the energy of 220 keV. So the detector is suitable for $\gamma$-ray detection with energy lower than 220 keV.

\begin{figure*}[!htb]
\includegraphics[width=.7\hsize]{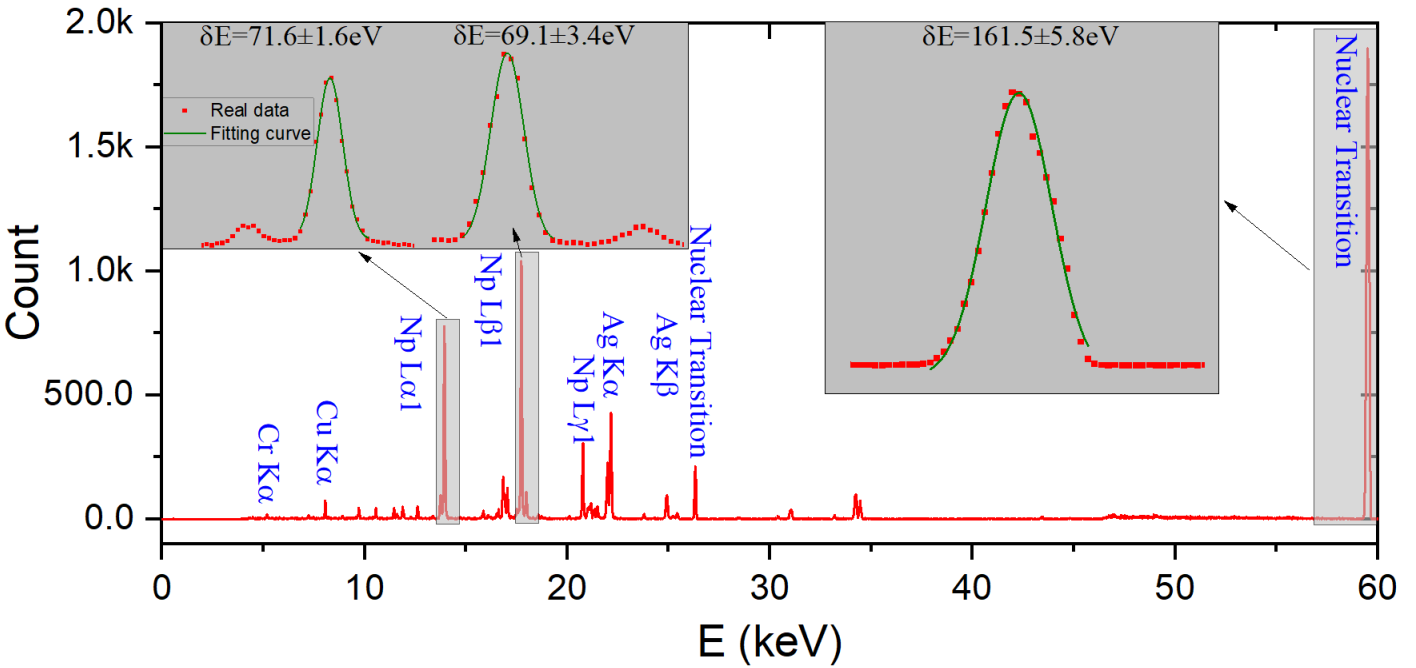}
\caption{The measured energy spectrum of Am241 can see spectral lines of $K_{\alpha}$ of copper, $L_{\alpha 1}$\&$L_{\beta 1}$ of NP, $K_{\alpha 1}$\&$K_{\alpha 2}$\&$K_{\beta}$ of silver and 26.3 keV\&59.5 keV nuclear transition lines of $\ce{^{241}Am}$.}
\label{fig12}
\end{figure*}

\subsection{Energy spectrum and resolution}\label{D}
The detected energy spectrum shown in Figure \ref{fig12} is obtained by accumulating the pulse eight of each signal calibrated using the least squre fit "quadratic" model inferred by data in Figure \ref{fig11}. By gaussian fitting, the detector has an energy resolution of 161.5 eV at 59.5 keV. However, at lower energies, such as 13.95 keV and 17.75 keV, the detector achieved energy resolutions of 71.6 eV and 69.1 eV. In theory, the energy resolution of TES $\gamma$-ray detector should be independent of the photon energy. Therefore the achieved results does not seem consistent with the theoretical expectation. The reason for this phenomenon requires subsequent more detailed analysis before conclusions can be drawn. In any case, a possible reasonable explanation of this inconsistency could be the following. It can be seen from the previous section that the detector is still well linear at 59.5 keV, so the problem is unlikely to be introduced by nonlinearity. As mentioned earlier, the higher the energy, the stronger the penetrating ability of the particle, so the thermalization path of the 59.5 keV photon in the absorption body is much more complicated than that of the 17.7 keV photon, which may be the main reason for this problem.

\section{Summary and Outlook} \label{sec.IV}
In summary, several sets of synchrotron radiation facilities and free electron laser facilities have been constructed or are under construction around Shanghaitech University, therefore, several beamline end stations are also being constructed simultaneously. Based on the energy spectrum measurement requirements of these end stations, Shanghaitech University has built a set of TES X-ray detectors for the X-ray band to meet the needs. Using this detector, the composition analysis of complex samples and the impurity analysis of high-purity samples have been carried out. On this basis, the R\&D team prepared a TES $\gamma$-ray detector for the energy band below 200 keV. After preliminary analysis and testing, the detector still has a quantum efficiency of 70\% near 100 keV, and has a single pixel count of 10 CPS. It is in the linear working region below 220 keV and has an energy resolution of 161.5 eV at 59.5 keV. However, at lower energies, such as 13.9 keV and 17.7 keV, the detector achieved energy resolutions of 71.6 eV and 69.1 eV. This resolution is two to three times higher than existing silicon drift detectors and HPGe detectors, and in the foreseeable future, higher energy resolution can be achieved with process optimization and more detailed signal analysis.

Although the energy resolution of this detector exceeds that of existing silicon drift detectors and HPGe detectors, its low count rate makes it unsuitable to replace these two detectors for general-purpose measurements. However, the detector's insensitivity to doping allows bury the sample inside the absorber of the detector. If the absorber is thick enough, the energy of the high energy particles will all be converted into a thermal signal and detected.

For X-ray and $\gamma$-ray metrology, this detector can be used for 4$\pi$ measurement of very low-activity radioactive substances. Long decay period radionuclides widely exist in nature, and their existence have a great impact on the development of radiopharmaceuticals and other materials. Due to its extremely low radioactivity, a 4$\pi$ measurement with high energy resolution must be taken to obtain sufficient sensitivity. The TES $\gamma$-ray detector can obtain an energy resolution better than HPGe on the premise of ensuring 4$\pi$ measurement, and is expected to improve the detection sensitivity by one to two orders of magnitude.

The measurement of dark matter is greatly affected by the purity of natural radioactive impurities in the liquid, and the beta decay energy spectrum data of natural radioactive substances is seriously lacking. The detector can accurately measure the low energy spectrum of beta ray from $\ce{^{226}Ra}$,$\ce{^{40}K}$ and other dark matter related nuclide. So, it could further depress the sensitivity of dark matter detection.

Neutron decay has been investigated as a possible pathway to explore of new physics, bond beta-decay of neutron(BOB) is an process that one neutron decay into a hydrogen and an anti-neutrino. Unlike normal neutron decay, it is a two-body process, so it provides a very elegant method to study neutrino helicities\cite{43-BOB-2014}. However, this process has not yet been observed so far for the very low branching ratio, the number can be as low as $10^{-6}$. BOB is a two-body process, the energy carried by the neutrino is fixed, and the remaining hydrogen atoms also have a fixed energy of about 325.7 eV. Due to the extremely strong penetrating power of neutrinos, the absorber can only intercept the energy of 325.7 eV, so the energy spectrum of this process in the TES X-ray detector appears as a single energy spectral line. Conventional neutron decay is a three-body process, and the sum of electron and proton energies will behave as a continuum. The energy resolution of the TES X-ray detector is in the order of eV, with the help of which, the BOB can be measured from the energy spectrum.



\begin{thebibliography} {99}
\bibitem{1-TES-2005} K.D.Irwin, G.C.Hilton, Transition-Edge Sensors. Cryogenic Particle Detection, Topics Appl. Phys.99. Chapter.3 (Springer, Berlin, Heidelberg), 63-150 (2005). \href{https://link.springer.com/chapter/10.1007\%2F10933596_3}{doi:10.1007/10933596\_3}
\bibitem{2-TESreview-2015} J.N.Ullom, D.A.Bennett, Review of superconducting transition-edge sensors for x-ray and $\gamma$-ray spectroscopy. Supercond. Sci. Technol. {\bf 28}, 084003 (2015). \href{https://iopscience.iop.org/article/10.1088/0953-2048/28/8/084003}{doi:10.1088/0953-2048/28/8/084003}
\bibitem{3-TESSRON-2021} L.Gottardi, K.Nagayashi, A Review of X-ray Microcalorimeters Based on Superconducting Transition Edge Sensors for Astrophysics and Particle Physics. Applied Sciences.{\bf11(9)}, 3793 (2021). \href{https://www.mdpi.com/2076-3417/11/9/3793}{doi:10.3390/app11093793}
\bibitem{4-Synchrotron-2006} S. Friedrich, Cryogenic X-ray detectors for synchrotron science. J. Synchrotron Rad. {\bf 13} 159 (2006) \href{https://journals.iucr.org/s/issues/2006/02/00/gf0004/index.html}{doi:10.1107/S090904950504197X}
\bibitem{5-Synchrotron-2015} J. Uhlig, W. B. Doriese, J. N. Ullom et al., High-resolution X-ray emission spectroscopy with transition-edge sensors: present performance and future potential. J. Synchrotron Rad. {\bf 22} 766 (2015) \href{https://journals.iucr.org/s/issues/2015/03/00/hf5280/index.html}{doi:10.1107/S1600577515004312 }
\bibitem{6-Synchrotron-2017} W. B. Doriese, J. N. Ullom, D. S. Swetz et al., A practical superconducting-microcalorimeter X-ray spectrometer for beamline and laboratory science. Rev. Sci. Instrum. {\bf 88} 053108 (2017) \href{https://aip.scitation.org/doi/10.1063/1.4983316}{doi:10.1063/1.4983316}
\bibitem{7-Ledge-2021} J. W. Fowler, G. C. O'Neil, J. N. Ullom et al., Absolute energies and emission line shapes of the L x-ray transitions of lanthanide metals. Metrologia {\bf 58}, 015016 (2021). \href{https://iopscience.iop.org/article/10.1088/1681-7575/abd28a}{doi:10.1088/1681-7575/abd28a}
\bibitem{8-JPTES-2020} S. Yamada, H. Tatsuno, T. Hashimoto et al., Coevolution of the Technology on Transition Edge Sensor Spectrometer and Its Application to Fundamental Science. J. Low Temp. Phys. {\bf 200} 418 (2020). \href{https://link.springer.com/article/10.1007/s10909-020-02441-2}{doi:10.1007/s10909-020-02441-2}
\bibitem{9-TRXAS-2013}	J. Uhlig, W.Fullagar, J.N. Ullom, et al., Table-Top Ultrafast X-Ray Microcalorimeter Spectrometry for Molecular Structure. Phys. Rev. Lett. {\bf 110}, 138302 (2013). \href{https://journals.aps.org/prl/abstract/10.1103/PhysRevLett.110.138302}{doi:10.1103/PhysRevLett.110.138302}
\bibitem{10-TRXES-2015} Y. I. Joe, G. C. O'Neil, J. N. Ullom et al., Observation of iron spin-states using tabletop x-ray emission spectroscopy and microcalorimeter sensors. J. Phys. B: At. Mol. Opt. Phys. {\bf 49}, 024003 (2015). \href{https://iopscience.iop.org/article/10.1088/0953-4075/49/2/024003}{doi:10.1088/0953-4075/49/2/024003}
\bibitem{11-TRXES-2017} G. C. O'Neil, L. M. Avila, J. N. Ullom et al., Ultrafast Time-Resolved X-ray Absorption Spectroscopy of Ferrioxalate Photolysis with a Laser Plasma X-ray Source and Microcalorimeter Array. J. Phys. Chem. Lett. {\bf 8(5)}, 1099-1104 (2017). \href{https://pubs.acs.org/doi/10.1021/acs.jpclett.7b00078}{doi:10.1021/acs.jpclett.7b00078}
\bibitem{12-TRXES-2016} L. M. Avila, G. C. O'Neil, J. N. Ullom et al., Ultrafast Time-Resolved Hard X-Ray Emission Spectroscopy on a Tabletop. Phys. Rev. X {\bf 6}, 031047 (2016).   \href{https://journals.aps.org/prx/abstract/10.1103/PhysRevX.6.031047}{doi:10.1103/PhysRevX.6.031047}
\bibitem{13-Hadron-2016} S. Okada, D. A. Bennett, J. Zmeskal et al., First application of superconducting transition-edge sensor microcalorimeters to hadronic atom X-ray spectroscopy. Prog. Theor. Exp. Phys. {\bf 09}, 091D01 (2016). \href{https://academic.oup.com/ptep/article/2016/9/091D01/2590795}{doi:10.1093/ptep/ptw130}
\bibitem{14-kaonic-2020}T. Hashimoto, D. A. Bennett, S. Yamada et al., Integration of a TES-based X-ray spectrometer in a kaonic atom experiment. J. Low Temp. Phys. {\bf 199} 1018 (2020).   \href{https://link.springer.com/article/10.1007\%2Fs10909-020-02434-1}{doi:10.1007/s10909-020-02434-1}
\bibitem{15-EBIT-2017} Y.Shen, J.Xiao, K.Yao et al., The status of the micro-calorimeter at Shanghai EBIT. Nucl. Instrum. Methods Phys. Res. Sect.  B {\bf 408}, 326–328 (2017). \href{https://doi.org/10.1016/j.nimb.2017.05.049}{doi:10.1016/j.nimb.2017.05.049}
\bibitem{16-MX-2020}	J. S. Adams, N. Bastidon, D. C. Goldfinger et al., First Operation of TES Microcalorimeters in Space with the Micro-X Sounding Rocket. J. Low Temp. Phys. {\bf 199} 1062-1071 (2020). \href{https://link.springer.com/article/10.1007\%2Fs10909-019-02293-5}{doi:10.1007/s10909-019-02293-5}
\bibitem{17-XIUF-2016} D. Barret, T. L. Trong, J. W. d. Herder et al., The Athena X-ray Integral Field Unit (X-IFU). Proc. of SPIE {\bf 9905}, 99052F-1 (2016). \href{https://www.spiedigitallibrary.org/conference-proceedings-of-spie/9905/1/The-Athena-X-ray-Integral-Field-Unit-X-IFU/10.1117/12.2232432.full?SSO=1}{doi:10.1117/12.2232432}
\bibitem{18-HUBS-2020} W. Cui, J. N. Bregman, M. P. Bruijn et al., HUBS: a dedicated hot circumgalactic medium explorer. Proc. of SPIE {\bf 11444}, 114442S (2020). \href{https://www.spiedigitallibrary.org/conference-proceedings-of-spie/11444/2560871/HUBS-a-dedicated-hot-circumgalactic-medium-explorer/10.1117/12.2560871.full}{doi:10.1117/12.2560871}
\bibitem{19-SEM-2020} M. H. Carpenter, M. P. Croce, Z. K. Baker et al., Hyperspectral X-ray Imaging with TES Detectors for Nanoscale Chemical Speciation Mapping. J. Low Temp. Phys. {\bf 200} 437-444 (2020). \href{https://link.springer.com/article/10.1007\%2Fs10909-020-02456-9}{doi:10.1007/s10909-020-02456-9}
\bibitem{25-NEXAS-2019} S. J. Lee,  C. J. Titus, K. D. Irwin et al., Soft X-ray spectroscopy with transition-edge sensors at Stanford Synchrotron Radiation Lightsource beamline 10-1.  Rev. Sci. Instrum. {\bf 90}, 113101 (2019). \href{https://aip.scitation.org/doi/10.1063/1.5119155}{doi:10.1063/1.5119155}
\bibitem{26-NEXAS-2019} S. F. Li, S. J. Lee, Y. J. Liu et al., Surface-to-Bulk Redox Coupling through Thermally Driven Li Redistribution in Li- and Mn-Rich Layered Cathode Materials. P. Am. Chem. Soc. {\bf 141(30)} 12079–12086 (2019). \href{https://pubs.acs.org/doi/10.1021/jacs.9b05349}{doi:10.1021/jacs.9b05349}
\bibitem{27-NEXAS-2017} C. J. Titus, M.L. Baker, D. Nordlund et al., L-edge spectroscopy of dilute, radiation-sensitive systems using a transition-edge-sensor array. J. Chem. Phys. {\bf 147}, 214201 (2017). \href{https://aip.scitation.org/doi/10.1063/1.5000755}{doi:10.1063/1.5000755}
\bibitem{28-RSXS-2020}	Y.I. Joe, Y. Z. Fang, P. Abbamonte et al., Resonant Soft X-Ray Scattering from Stripe-Ordered $\rm{La_{2-x}Ba_xCuO_4}$ Detected by a Transition-Edge Sensor Array Detector. Phys. Rev. Applied. {\bf 13}, 034026 (2020). \href{https://journals.aps.org/prapplied/abstract/10.1103/PhysRevApplied.13.034026}{doi:10.1103/PhysRevApplied.13.034026}
\bibitem{29-EBIT-2014} L. Gabriele, B. Martinez, J.Ullom et al., The transition-edge EBIT microcalorimeter spectrometer. Proc. of SPIE {\bf 9144}, 91443U (2014). \href{https://www.spiedigitallibrary.org/conference-proceedings-of-spie/9144/1/The-transition-edge-EBIT-microcalorimeter-spectrometer/10.1117/12.2055568.full?SSO=1}{doi:10.1117/12.2055568}
\bibitem{30-EBIT-2009} G. V. Browna, J. S. Adamsb, P. Beiersdorfer et al., Laboratory Astrophysics, QED, and other Measurements using the EBIT Calorimeter Spectrometer at LLNL. AIP Conference Proceedings {\bf 1185}, 446 (2009). \href{https://aip.scitation.org/doi/abs/10.1063/1.3292374}{doi:10.1063/1.3292374}
\bibitem{31-EBIT-2019}  P. Szypryt,  G. C. O'Neil, E. Takacs et al., A transition-edge sensor-based x-ray spectrometer for the study of highly charged ions at the National Institute of Standards and Technology electron beam ion trap.  Rev. Sci. Instrum. {\bf 90}, 123107 (2019). \href{https://aip.scitation.org/doi/10.1063/1.5116717}{doi:10.1063/1.5116717} \href{https://www.spiedigitallibrary.org/conference-proceedings-of-spie/11838/118380X/NIST-microcalorimeter-arrays-for-the-hard-x-ray-and-\%ce\%b3/10.1117/12.2594652.full}{doi:10.1117/12.2594652}
\bibitem{33-DR-2014} K.Maehataa, T.Harab, T.Ito et al., A dry 3He–4He dilution refrigerator for a transition edge sensor microcalorimeter spectrometer system mounted on a transmission electron microscope. Cryogenics {\bf 61}, 86-91 (2014). \href{https://www.sciencedirect.com/science/article/pii/S0011227514000538?via\%3Dihub}{doi:10.1016/j.cryogenics.2014.03.002}
\bibitem{34-SEM-2021} P. Szypryt,  D. A. Bennett, W.J. Boone et al., Design of a 3000-Pixel Transition-Edge Sensor X-Ray Spectrometer for Microcircuit Tomography. EEE Transactions on Applied Superconductivity {\bf 31(5)}, 2100405 (2021). \href{https://ieeexplore.ieee.org/document/9328316}{doi: 10.1109/TASC.2021.3052723}
\bibitem{20-NuCl-2019} A. Yamaguchi, H. Muramatsu, K. Mitsuda et al., Energy of the 229Th Nuclear Clock Isomer Determined by Absolute $\gamma$-ray Energy Difference. Phys. Rev. lett. {\bf 123}, 222501 (2019). \href{https://journals.aps.org/prl/abstract/10.1103/PhysRevLett.123.222501}{doi:10.1103/PhysRevLett.123.222501}
\bibitem{21-securaty-2009} M. W. Rabin, National and International Security Applications of Cryogenic Detector Mostly Nuclear Safeguards. AIP Conference Proceedings {\bf 1185}, 725-732 (2009). \href{https://aip.scitation.org/doi/10.1063/1.3292444}{doi:10.1063/1.3292444}
\bibitem{22-256gamma-2015} R.Winklera, A.S.Hoover, J.N.Ullom et al., 256-pixel microcalorimeter array for high-resolution $\gamma$-ray spectroscopy of mixed-actinide materials. Nucl. Instrum. Methods Phys. Res. Sect. A {\bf 770}, 203-210 (2015). \href{https://doi.org/10.1016/j.nima.2014.09.049}{doi:10.1016/j.nima.2014.09.049}
\bibitem{35-securaty-2015} R.Winkler, A.S.Hoover, J.N.Ullomb et al., 256-pixel microcalorimeter array for high-resolution $\gamma$-ray spectroscopy of mixed-actinide materials. Nucl. Instrum. Methods Phys. Res. Sect. A {\bf 770}, 203-210 (2015).  \href{https://doi.org/10.1016/j.nima.2014.09.049}{doi: 10.1016/j.nima.2014.09.049}
\bibitem{36-Physica-2021} S.Zhang, W.Cui, Z.Liu et al., Development of basic theory and application of cryogenic X-ray spectrometer in light sources and X-ray satellite. Acta Phys. Sin {\bf 70}, 180702 (2021).   \href{https://wulixb.iphy.ac.cn/article/doi/10.7498/aps.70.20210350}{doi: 10.7498/aps.70.20210350}
\bibitem{37-microcalorimeter-1984} S.H.Moseley, J.C.Mather, D.McCammon et al., Thermal detectors as X-ray spectrometers. Journal of Applied Physics. {\bf 56}, 1257 (1984). \href{https://aip.scitation.org/doi/10.1063/1.334129}{doi: 10.1063/1.334129}
\bibitem{38-MicroCal-2005} D. McCammon, Thermal Equilibrium Calorimeters – An Introduction. Cryogenic Particle Detection, Topics Appl. Phys.99. Chapter.1 (Springer, Berlin, Heidelberg), 1-34 (2005). \href{https://doi.org/10.1007/10933596_1}{doi:10.1007/10933596\_1}
\bibitem{39-SemMC-2005} D. McCammon, Semiconductor Thermistors. Cryogenic Particle Detection, Topics Appl. Phys.99. Chapter.2 (Springer, Berlin, Heidelberg), 35-62 (2005). \href{https://doi.org/10.1007/10933596_2}{doi:10.1007/10933596\_2}
\bibitem{40-MMC-2005} . FleischmannC. EnssG.M. Seidel, Metallic Magnetic Calorimeters. Cryogenic Particle Detection, Topics Appl. Phys.99. Chapter.4 (Springer, Berlin, Heidelberg), 151-216 (2005). \href{https://doi.org/10.1007/10933596_4}{doi:10.1007/10933596\_4}
\bibitem{41-neotrino-2016} M. P. Croce, M. W. Rabin, V. Mocko et al., Development of Holmium-163 Electron-Capture Spectroscopy with Transition-Edge Sensors. J. Low Temp. Phys. {\bf 184} 958–968 (2016). \href{https://link.springer.com/article/10.1007/s10909-015-1451-2}{doi: 10.1007/s10909-015-1451-2}
\bibitem{42-SQUID-201} J.D. Eschweiler, A superconducting microcalorimeter for low-flux detection of near-infrared single photons. PHD. Dissertation (Hamburg: University of Hamburg) ,(2014). \href{https://doi:10.3204/DESY-THESIS-2014-016}{doi: 10.3204/DESY-THESIS-2014-016}
\bibitem{43-BOB-2014} J.McAndrew, S. Paul et al., Bound Beta-decay of the Free Neutron: BoB. Physics Procedia {\bf 51}, 37 (2014). \href{https://www.sciencedirect.com/science/article/pii/S1875389213006974}{doi: 10.1016/j.phpro.2013.12.009}
\end{thebibliography}
\end{document}